# On the pseudogap and doping-dependent magnetic properties of La$_{2-x}$Sr$_x$Cu$_{1-y}$Zn$_y$O$_4$

R. S. Islam[a,b], J. R. Cooper[b], J. W. Loram[b], and S. H. Naqib[a,b*]

[a] Department of Physics, Rajshahi University, Raj-6205, Bangladesh

[b] IRC in Superconductivity and Department of Physics, University of Cambridge, Madingley Road, Cambridge CB3 0HE, UK

**Abstract**

The effects of planar hole content, p (= x), on the uniform (q = 0) magnetic susceptibility, $\chi(T)$, of La$_{2-x}$Sr$_x$Cu$_{1-y}$Zn$_y$O$_4$ were investigated over a wide range of Sr (x) and Zn (y) contents. A strongly p-dependent Zn-induced magnetic behavior was observed. The apparent Zn-induced magnetic moment is larger in underdoped La$_{2-x}$Sr$_x$Cu$_{1-y}$Zn$_y$O$_4$ and it decreases quite sharply around p ~ 0.19. It does not change much for further overdoping. This indicates a possible role of the pseudogap on the Zn induced magnetic behavior, as there is growing evidence that pseudogap vanishes quite abruptly at p ~ 0.19.

*Keywords*: Pseudogap; Magnetic propertires of L214; Zn substituted La214

## 1. Introduction

Zn$^{2+}$ is nonmagnetic due to its 3d$^{10}$ electronic configuration, but various experiments have indicated that Zn gives rise to a local magnetic moment-like feature, possibly on the four nearest neighbour Cu sites in the CuO$_2$ planes of high-T$_c$ superconductors [1]. This magnetic behaviour is evident from the appearance of a Curie-like term in the bulk magnetic susceptibility of Zn-doped samples [2]. A complete understanding of this effect is still lacking [3]. Here, we report some systematic studies of the effects of planar hole content, p, on the uniform magnetic susceptibility, $\chi(T)$, of polycrystalline La$_{2-x}$Sr$_x$Cu$_{1-y}$Zn$_y$O$_4$ samples. From the analysis of the $\chi(T)$ data, we have found the Zn-induced increase in $\chi(T)$ at low T to be strongly dependent on p and consequently possibly on the presence of the pseudogap (PG).

## 2. Samples and experimental results

Polycrystalline single-phase samples of La$_{2-x}$Sr$_x$Cu$_{1-y}$Zn$_y$O$_4$ were synthesized (with x = 0.09, 0.15, 0.19, 0.22, 0.27 and y = 0.0, 0.005, 0.01, 0.015, 0.02, and 0.024, for each value of x) using the solid-state reaction method. Samples were characterized by X-ray diffraction, thermopower, and AC susceptibility measurements, details can be found in ref. [4]. $\chi(T)$ was measured using a *model 1822 Quantum Design* SQUID magnetometer (in a DC field of 5 Tesla). Representative $\chi(T)$ data are shown in Fig. 1. These plots show that (i) a Curie-like enhancement appears at low T and increases with Zn, (ii) this enhancement decreases with increasing x (= p for LSCO), and (iii) for samples with x ≥ 0.19, the enhancement is similar in magnitude to that of x = 0.19. Therefore, the effect of Zn on $\chi(T)$ is greater for underdoped samples.

## 3. Analysis of data

Data have been analyzed using $\chi(T)T$ plots (see Fig. 2). For a T- region where the plots are parallel,

$$\chi(T)T = \chi_h(T)T + C \qquad (1)$$

where, $\chi_h(T)$ is the susceptibility of the host material (Zn-free LSCO), which in the first approximation, is assumed not to change significantly with Zn, and C, the Curie constant, is given by:

$$C = (N_A n \mu_B^2 P_{eff}^2)/3k_B \qquad (2)$$

here, N$_A$ is Avogadro's number, n is the Zn content per mole, $\mu_B$ is the Bohr magneton, P$_{eff}$ is the effective magneton number, and k$_B$ is Boltzmann's constant. In practice, $\chi(T)T$ plots for different values of y and the same x are not exactly parallel at all T. We have selected T-regions above $T_c$ (typically 60K - 110K) where the $\chi(T)T$ were linear (see Fig. 2) and parallel for all values of y. We then obtained values of C by linear regression. From the slopes of C versus y, $P_{eff}^2$/Zn (the Zn-induced "*effective moment*" in units of $\mu_B^2$) was calculated using Eqn. (2) [4] and is plotted versus x (= p) in Fig. 3. $P_{eff}^2$/Zn decreases with increasing p, as found by previous studies [3,5,6] for Zn-doped LSCO and YBCO. It is worth noting that in these works slightly different analyses were used, but fairly consistent values for $P_{eff}^2$/Zn were obtained in all studies including the present one.

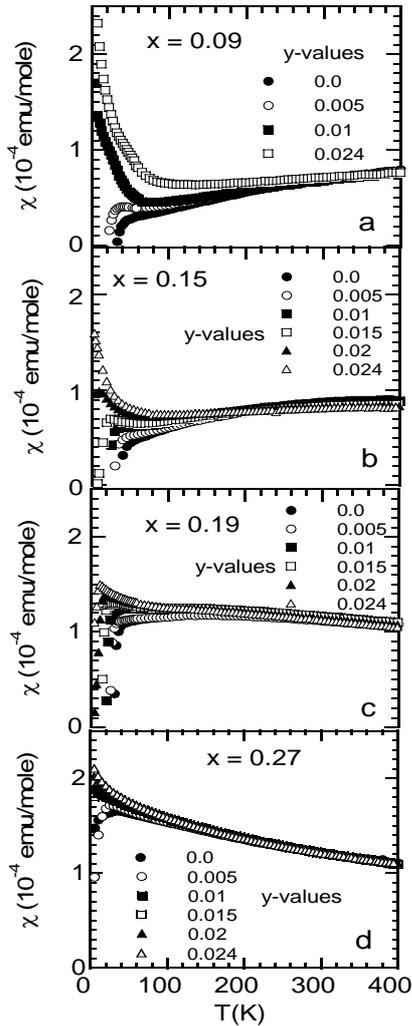

Fig. 1. $\chi(T)$ for some representative $La_{2-x}Sr_xCu_{1-y}Zn_yO_4$ samples.

### 4. Discussion and conclusions

Fig. 3 shows one of the initial findings of the present study. It is seen that $P_{eff}^2$/Zn falls quite sharply until p ~ 0.19 and that further overdoping does not change the Zn-induced magnetic behavior significantly. It is also worth noticing that even near optimum p ~ 0.15, where $T_c$ is maximum, $P_{eff}^2$/Zn is quite large. There have been suggestions based on experimental observations that the PG vanishes at p ~ 0.19 for cuprates [7]. A detailed analysis of our data is in progress in order to clarify the role played by the PG on the Zn-induced magnetic behavior in cuprates.

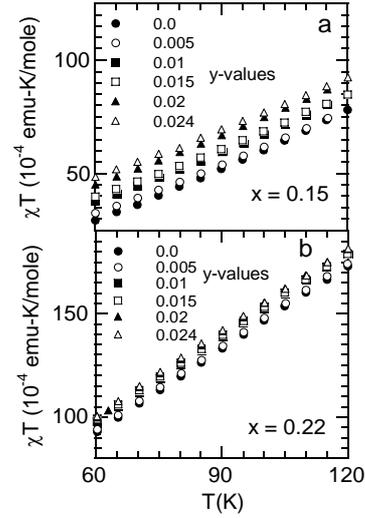

Fig. 2. $\chi(T)T$ for some representative $La_{2-x}Sr_xCu_{1-y}Zn_yO_4$ samples.

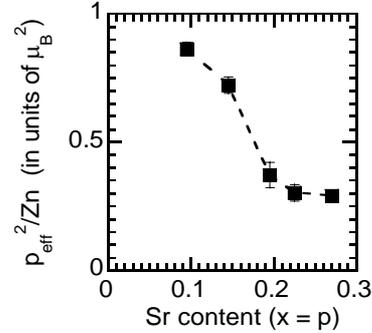

Fig. 3. $P_{eff}^2$/Zn versus hole content for $La_{2-x}Sr_xCu_{1-y}Zn_yO_4$ samples.

### Acknowledgments

The authors thank Prof. J. L. Tallon for helpful comments and suggestions. SHN thanks the members of the Quantum Matter group, Department of Physics, University of Cambridge, UK, for their hospitality.